\documentclass{article}
\usepackage{spadre2008}
\usepackage{graphicx}
\frompage{000} \topage{000}                                              
\def\rightmark{Mach-like Correlations from AdS/CFT Jets}
\def\leftmark{J. Noronha and M. Gyulassy}

\title{Mach-like Angular Correlations Arise Only
From the Head Zone of AdS/CFT String Jets}
\authors{
{Jorge Noronha $^{a}$, Miklos Gyulassy $^{b}$%
}\\[2.812mm]
{\normalsize
\hspace*{-8pt} Department of Physics, Columbia University, 538 West 120$^{th}$ Street,\\ New York, NY 10027, USA \\
}}

\abstract{We show that Mach-like azimuthal correlations of associated hadrons
in the wake of heavy quark jets modeled in the AdS/CFT string drag picture originate only from the near-quark ``Head'' region. The contributions from the far zones lead only
to a yield peaked in the opposite direction with respect to the trigger jet. }

\keyword{AdS/CFT, strongly coupled quark-gluon plasma, heavy quark jets, Mach cones.}
\PACS{25.75.-q, 11.25.Tq, 13.87.-a}

\begin{document}

\maketitle
\setcounter{page}{1}

\section{Introduction}\label{intro}
The observation of Mach cone-like correlations between tagged jets and associated hadrons \cite{Adler:2005ee} has been interpreted as providing evidence for the very fast relaxation time of energy-momentum perturbations in the quark-gluon plasma \cite{Stoecker:2004qu,shuryakcone,heinzcone,renk,Betz:2008js}. While perturbative quantum chromodynamics (pQCD)  provides a very good description of the attenuation of light quark and gluon jets observed at the relativistic heavy ion collider (RHIC) \cite{Gyulassy:2003mc,GLV}, the recent non-photonic single electron data \cite{Adler:2005xv} have proven to be more complicated to be fully accounted for in pQCD-based models \cite{Djordjevic:2005db}. The small heavy quark diffusion coefficient that seems to be required to fit the data is in the same ballpark as that derived \cite{diffusionAdS} using the anti-de Sitter/Conformal Field Theory (AdS/CFT) correspondence \cite{maldacena}.

The first attempts towards understanding jet quenching through AdS/CFT were proposed in \cite{diffusionAdS,Liu:2006ug,Herzog:2006gh} in the context of the duality between $\mathcal{N}=4$ supersymmetric Yang-Mills (SYM) plasmas and type IIB string theory in AdS$_{5}\otimes S_{5}$ in the supergravity approximation (where $N_c \rightarrow\infty$ and $\lambda\equiv g^2 N_c\gg 1$). The trailing string solution derived in \cite{Herzog:2006gh} provided a very thorough description of heavy quark energy loss in strongly-coupled $\mathcal{N}=4$ SYM plasmas. Basically, in this model one considers the effects created by an external probe, a heavy quark moving at a constant velocity $v$, on a static $\mathcal{N}=4$ SYM plasma at temperature $T_0$. We remark that theories with supergravity dual
descriptions do not display true jet-like structures because, in average, their particle energy distributions are spherically symmetric \cite{Hofman:2008ar} (see also \cite{spherical}).
However, in steady-state solutions such as the trailing string, the asymmetry is already included in the initial state (for instance, our jet is defined to move along the $X_1$ axis).

Highly energetic back-to-back jets produced in the early stages of
the collision can traverse the evolving strongly-coupled quark-gluon plasma (sQGP) and deposit energy along
their path. In the case where one of the jets is produced near the
surface (trigger jet), the other supersonic away side jet can move
through most of the plasma and generate a Mach cone, which is expected to
lead to an enhancement of particles emitted at the Mach angle
\cite{Stoecker:2004qu,shuryakcone}. In the hard momentum region where jets are produced, the relevant scale is much larger than the temperature, $Q\gg T_{0}$, and pQCD is certainly the correct description. In the soft part of the process ($Q\sim T_0$)  \cite{Mueller:2008zt}, we expect that AdS/CFT can shed some light into the complicated (and possibly strongly-coupled) non-equilibrium dynamics present in the interaction between, for instance, heavy quark jets and the underlying sQGP. In fact, it has been shown that Mach cones are produced when a supersonic heavy quark (as defined by \cite{Herzog:2006gh})
that moves through a static strongly-coupled $\mathcal{N}=4$ SYM plasma
\cite{Chesler:2007an,gubsermach}. Direct tests of the AdS/CFT string drag model using the nuclear modification factor, $R_{AA}(p_T)$, have been proposed for at the LHC
\cite{Horowitz:2007su}.

In this paper we investigate whether the 2-particle angular correlations associated with the supersonic AdS/CFT heavy quark jet display a double-peak structure. We take the medium to be a static strongly-coupled $\mathcal{N}=4$ SYM thermal plasma and the AdS/CFT description of heavy quark energy loss derived in \cite{Herzog:2006gh}. In our scenario the supersonic heavy quark travels through the plasma until it eventually punches through the medium and fragments in vacuum (we do not attempt to describe what happens to the quark after it left the medium). We use the numerical results of Ref.\ \cite{gubsermach} to show that an extrapolation to $N_c=3$ and large $\lambda$ leads to Mach cone correlations which strongly depend on the non-equilibrium region defined by the {\it head} of the jet.

\section{Cooper-Frye Freeze-out}\label{techno}


The total stress tensor $T^{\mu\nu}$ ($\mu,\nu=0,\ldots,3$) of the system given by the plasma+heavy quark has been computed in the case of a static background
\cite{Friess:2006fk}. When $N_c \rightarrow \infty$ the disturbances in the
energy-momentum tensor $\delta T^{\mu\nu}$ caused by the trailing string are
contributions of order $\sqrt{\lambda}/N_{c}^{2}$ with respect to
the background. A direct consequence of that is that linearized
first-order Navier-Stokes hydrodynamics is predicted to be a very good description
of the quark's wake down to distances of $\sim 1/T$ away from the
heavy quark \cite{Chesler:2007sv,Noronha:2007xe} (see also
\cite{Gubser:2008vz}). Once $T^{\mu\nu}(X)$ is determined, the flow
velocity $U^{\mu}(X)=\left(\sqrt{1+\vec{U}^2},\vec{U}\right)$ can be
obtained by boosting the system to its local rest frame and solving
the corresponding coupled nonlinear equations for the flow spatial
components \cite{Noronha:2007xe}. One can show that when
$N_c \rightarrow \infty$ the complicated equations that determine
the flow have a simple solution $U^{i}=T^{0i}/\left(4 p_0\right)$,
where $p_0=\pi^2(N_c^2-1)T_0^4/8$ is the dominant contribution for the pressure \cite{Chesler:2007sv,Noronha:2007xe}. Thus, since
$T^{0i}$ is only proportional to $\sqrt{\lambda}$, we see that
$U\sim \mathcal{O}(\sqrt{\lambda}/N_{c}^2)$, which is a very small
quantity when $N_{c}\rightarrow \infty$ and $g\to 0$. The
local temperature of the plasma will be, in general,
$T(X)=T_{0}+\Delta T(X)$, where $\Delta T(X)/T_0$ is also at least
of $\mathcal{O}(\sqrt{\lambda}/N_{c}^2)$. Using the approximation
$U^{i}=T^{0i}/(4p_0)$, one obtains that
\begin{equation}
\ T(X)=\left[\frac{8 \,T^{00}(X)}{3(N_{c}^2-1)\pi^2}\right]^{1/4}
\end{equation}
to leading order in $1/N_c$, where $T^{00}(X)=\varepsilon_0+\delta T^{00}(X)$ and $\varepsilon_0=3p_0$ is the background energy density of the plasma \cite{Gubser:1996de}.

We assume, as in \cite{shuryakcone,heinzcone,renk,Betz:2008js}, that the
energy and momentum deposited by the jet are converted into observable particles via the  Cooper-Frye (CF) freeze-out scheme \cite{Cooper:1974mv} in which the angular distribution of associated particles with 4-momentum
$P^{\mu}=(p_{T},p_{T}\cos (\pi-\phi),p_{T}\sin
(\pi-\phi),0)$\footnote{In this paper we focus on 2-particle
correlations only. Moreover, due to conformal invariance, the
emitted particles are nearly massless, i.e.,
$\sqrt{p_{T}^2+m^2}/T_{0}\rightarrow p_{T}/T_{0}$.} at mid rapidity
$y=0$ that is induced
by the jet is
\begin{equation}
\frac{dN}{dyd\phi}\Big
|_{y=0}=\int_{p_{T}^{i}}^{p_T^{f}}dp_{T}\,p_{T}\int_{\Sigma}d\Sigma_{\mu}P^{\mu}\left[f(U^{\mu},P^{\mu},T)-f_{eq}\right]
\label{cooperfrye}
\end{equation}
where $p_T$ is the transverse momentum, $p_T^{i,f}$ define the relevant $p_T$ bin, $\Sigma(X)$ is the freeze-out hypersurface, and $f=f_0+\delta f$ is the distribution
function which is the sum of an ideal Fermi-Dirac/Bose-Einstein
local distributions $f_0=1/\left(e^{-U^{\mu}P_{\mu}/T(X)}\pm 1\right)$ and $\delta f$ represents viscous contributions to the freeze-out.
This term is important for the description of the quantities
associated with the hydrodynamic properties of the RHIC plasma (see
for instance the recent study in \cite{Dusling:2007gi}). However, in
the supergravity limit $\delta f$ is only subleading in $1/N_c$ and
can then be neglected in our analysis. Note that we remove the yield
coming from the background by subtracting $f_{eq}\equiv f
|_{U^{\mu}=0,T=T_0}$. Moreover, we use here the Boltzmann limit
where $f_0 \left(1\pm f_0\right)\to f_{0}$. The hypersurface $\Sigma(X)$ will be a complicated function in the case of an expanding background. However, in the static case the natural CF
prescription is to use an isochronous freeze-out where
$d\Sigma_{\mu}=(dV,0,0,0)$ and $V$ is the 3-dimensional volume
\cite{shuryakcone}. Because of the axial symmetry with
respect to the jet axis, we assume that $U^{\mu}(X)=U^{\mu}(X_1,
X_{\perp})=(U_1,U_{\perp},0)$ where $X_1=X_{\parallel}-vt$ is the
co-moving coordinate along the jet direction and $X_{\perp}$
represents the transverse coordinates. Additionally,
$dV=dX_{\parallel}dX_{\perp}d\varphi\, X_{\perp}$ and $\vec{p}\cdot
\vec{U}=p_{T}U_{1}\cos (\pi-\phi)+p_{T}U_{\perp}\sin (\pi-\phi) \cos
\varphi$.

\section{Results}
\begin{figure}[htb]
\vspace*{1cm}
\insertplot{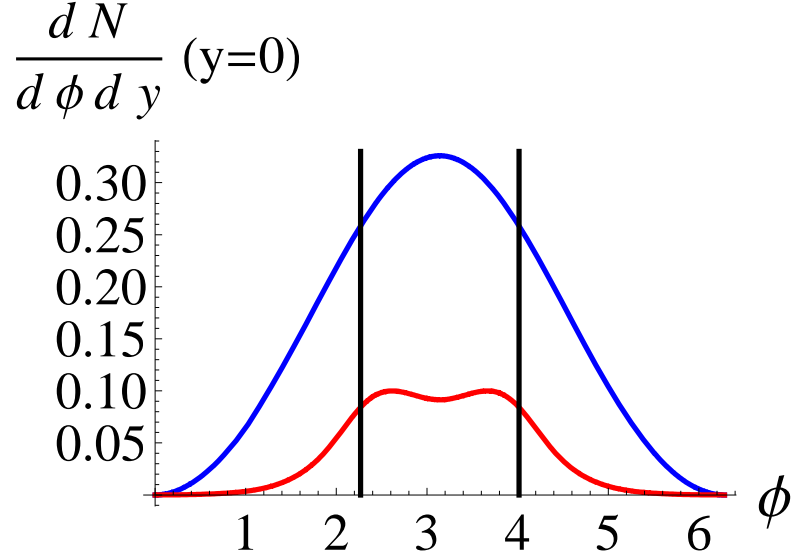}
\caption{Azimuthal correlations in AdS/CFT drag picture
for the case where $v=0.9$, $\lambda=5.5$, $N_c=3$ for  $3<p_T/ (\pi T_0)<4$.
The isochronous Cooper-Frye volume used for blue curve excludes the ``head'' zone
 where $\delta T^{00}(x)>0.3\, \varepsilon_0$. A Mach like red curve results
only from the head zone. The vertical black lines are located at $\pi\pm\phi_{M}$, where $\phi_M=\cos^{-1}c_s/v$. } \label{fig1}
\end{figure}

We solved Eq.\ (\ref{cooperfrye}) for $N_c=3$, $v=0.9$, and $\lambda=5.5$ \cite{gubserlambda} in a static
background using the numerical results for the $T^{00}$ and $T^{0i}$ components of
the stress tensor computed by Gubser, Pufu, and Yarom in Ref.\ \cite{gubsermach}.  In this solution, the infinitely massive heavy quark
has been moving since $t \rightarrow -\infty$ and can be found at
the origin at $t=0$. Our total CF volume is defined by $-14<X_1\,(\pi T_0)<1$, $0<X_{\perp}\,(\pi T_0)<14$, and $\varphi\in [0,2\pi]$. Note that the temperature is the only energy scale in the problem and that our system is always in the deconfined phase. Moreover, we remark that the leading order expressions we use for $U^{\mu}$(X) and $T(X)$ become less reliable in the vicinity of the heavy quark (see \cite{Yarom:2007ni,Gubser:2007nd} for a detailed study of the stress tensor in the near-quark region) because there the flow and temperature gradients become large. In fact, in the head region the relative magnitude of disturbance in the energy density $\xi(X)=\delta T_{00}(X)/\varepsilon_0$, where $\varepsilon_0=3p_0$, approaches the limit $\xi\sim 1$. Here we define the head of the jet as the volume where $\xi>0.3$, which roughly corresponds to the region between $-1<X_1\,(\pi T_0)<1$ and $0<X_{\perp}\,(\pi T_0)<2$. We show our results for the angular correlations in Fig.\ \ref{fig1} when $3<p_T/ (\pi T_0)<4$. The blue curve was computed excluding the near-quark head region from the CF volume. On the other hand, for the red curve we only included the head region where $\xi>0.3$. Note that the red curve displays a double-peak structure while the blue curve does not. The peaks appear at angles similar to the Mach cone angle $\pi\pm\phi_{M}$, where $\phi=\cos^{-1}c_s/v$ and $c_s=1/\sqrt{3}$ is the speed of sound.

\section{Conclusions}\label{concl}

In this work we studied the 2-particle angular correlations associated with AdS/CFT heavy quark jets. We showed that only the yield that comes from the non-equilibrium head region displays a double-peak structure. In fact, once the head region is excluded from the CF volume one obtains only a peak centered at $\phi=\pi$, which means that most of the produced particles are emitted in the opposite direction with respect to the trigger jet. A more detailed analysis of the interplay between the different regions involved in the CF freeze-out of the wake created by AdS/CFT heavy quark jets will presented elsewhere \cite{future}.

\section*{Acknowledgments}

We thank S. Gubser, S. Pufu, and A. Yarom for providing
their numerical results for the stress tensor and B. Betz and G. Torrieri for useful discussions. J.N. and M.G. acknowledge partial
support from DOE under Grant No. DE-FG02-93ER40764.

\section*{Note(s)}
\begin{notes}
\item[a]
E-mail: noronha@phys.columbia.edu
\item[b]
E-mail: gyulassy@phys.columbia.edu
\end{notes}

\vfill\eject
\end{document}